\documentclass[sn-mathphys,Numbered]{sn-jnl}

\usepackage{graphicx}%
\usepackage{multirow}%
\usepackage{amsmath,amssymb,amsfonts}%
\usepackage{amsthm}%
\usepackage{mathrsfs}%
\usepackage[title]{appendix}%
\usepackage{xcolor}%
\usepackage{textcomp}%
\usepackage{manyfoot}%
\usepackage{booktabs}%
\usepackage{algorithm}%
\usepackage{algorithmicx}%
\usepackage{algpseudocode}%
\usepackage{listings}%
\usepackage{lineno}
\usepackage{hyperref}

\theoremstyle{thmstyleone}%
\theoremstyle{thmstyletwo}%

\theoremstyle{thmstylethree}%

\begin{document}

\title[Article Title]{End-to-End Modeling of the TDM Readout System for CMB-S4}


\author*[1, 2]{D.~C.~Goldfinger}\email{dgoldfin@stanford.edu}
\author[1, 2]{Z.~Ahmed} 
\author[3]{D.~R.~Barron} 
\author[4]{W.~B.~Doriese} 
\author[4, 5]{M.~Durkin} 
\author[6]{J.~P.~Filippini} 
\author[2]{G.~Haller} 
\author[1, 2]{S.~W.~Henderson} 
\author[2]{R.~Herbst}
\author[4]{J.~Hubmayr} 
\author[1, 2]{K.~Irwin} 
\author[2]{B.~Reese}
\author[2]{L.~Sapozhnikov}
\author[1, 7]{K.~L.~Thompson}
\author[4]{J.~Ullom} 
\author[4]{M.~R.~Vissers} 

\affil*[1]{Kavli Institute for Particle Astrophysics and Cosmology, Menlo Park, CA 94025}
\affil[2]{SLAC National Accelerator Laboratory, Menlo Park, CA 94025}
\affil[3]{Department of Physics and Astronomy, University of New Mexico, Albuquerque, NM, USA}
\affil[4]{National Institute of Standards and Technology, 325 Broadway, Boulder, CO 80305 USA}
\affil[5]{Department of Physics, University of Colorado, Boulder, CO 80309}
\affil[6]{Department of Physics, University of Illinois at Urbana-Champaign, Urbana, IL 61801}
\affil[7]{Department of Physics, Stanford University, Stanford, CA 94305}

\abstract{The CMB-S4 experiment is developing next-generation ground-based microwave telescopes to observe the Cosmic Microwave Background with unprecedented sensitivity.  This will require an order of magnitude increase in the 100 mK detector count, which in turn increases the demands on the readout system.  The CMB-S4 readout will use time division multiplexing (TDM), taking advantage of faster switches and amplifiers in order to achieve an increased multiplexing factor.  To facilitate the design of the new readout system, we have developed a model that predicts the bandwidth and noise performance of this circuitry and its interconnections.  This is then used to set requirements on individual components in order to meet the performance necessary for the full system. We present an overview of this model and compare the model results to the performance of both legacy and prototype readout hardware.}




\maketitle

\section{Introduction}\label{sec1}

CMB-S4 will be a next-generation ground-based observatory, with telescopes located at the South Pole and in the Atacama Desert, to make polarized measurements of the Cosmic Microwave Background (CMB) across the southern sky with unprecedented sensitivity \cite{S4ScienceBook}.  It will access a range of science topics including testing the predictions of cosmic inflation, searching for evidence of additional light relic particles, characterizing large scale phenomena and searching for transient phenomena, among others.  Large Aperture Telescopes (LATs) will provide arcminute resolution maps of the southern sky, while Small Aperture Telescopes (SATs) provide smaller and deeper degree-scale maps with exquisite systematic errors.  Telescopes will observe at several CMB-adjacent wavelengths in order to facilitate component separation between the CMB and galactic foreground signals.

This experiment uses arrays of DC-voltage biased Transition Edge Sensor (TES) bolometers, with signals amplified by DC SQUIDs and read out through Time-Division Multiplexing (TDM) \cite{Barron22}.  CMB-S4 will have an order of magnitude increase in the number of detectors per focal plane as compared with the previous generation of CMB telescopes, targeting $\sim$130,000 detectors on the densest focal planes and $\sim$500,000 detectors overall.  In order to accommodate this increase, we are seeking to increase the multiplexing factor of the readout to reduce cost and integration complexity, by incorporating new approaches and technological developments.

\begin{figure}[h!]%
\centering
\includegraphics[width=0.95\textwidth]{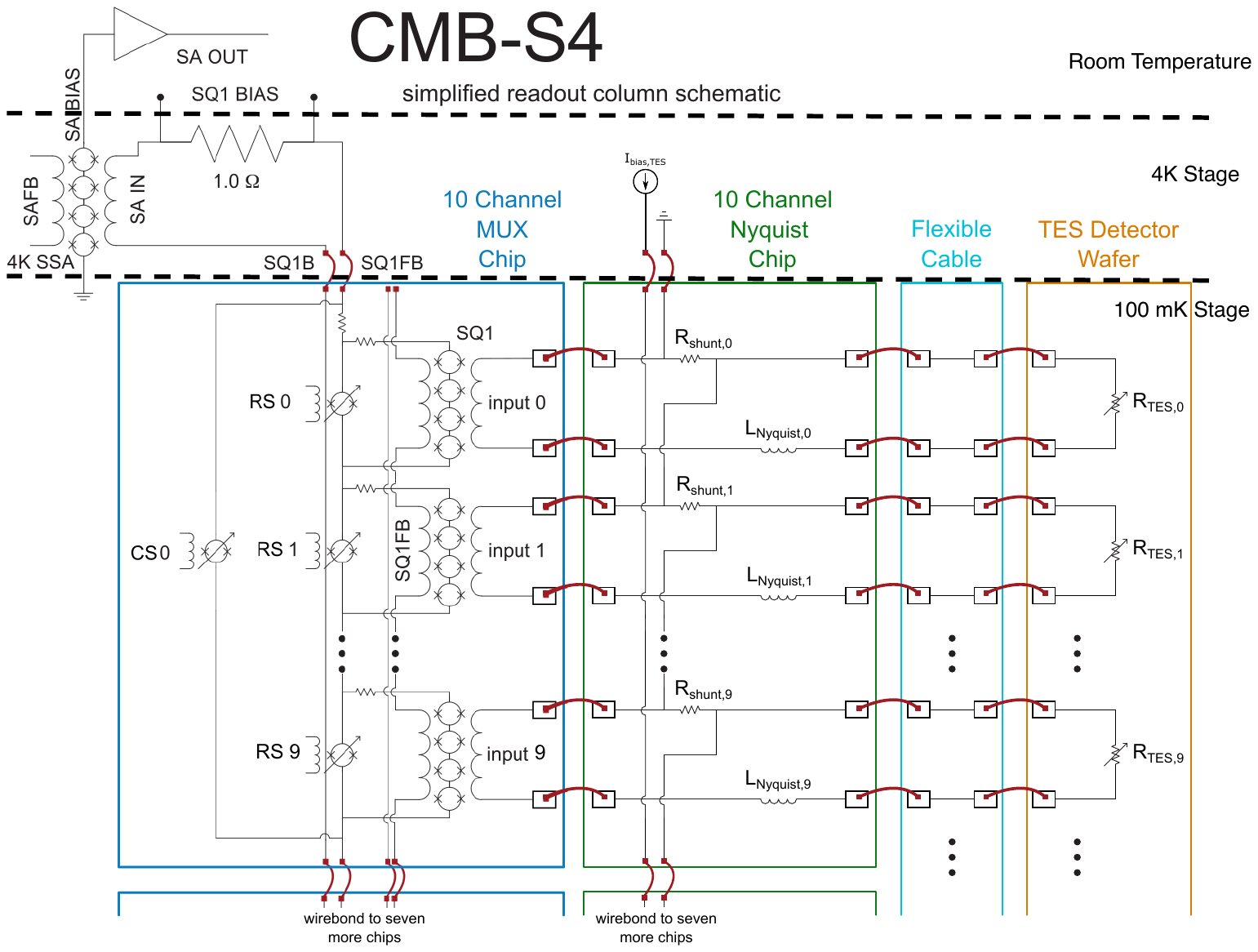}
\caption{Circuit diagram for the CMB-S4 readout electronics.  100 mK components are identified on the relevant chips, with simplified inputs for the wiring.  The column signals are depicted as coming from above, while the switches used for the rows leave their traces off to the left. \cite{Barron22}}\label{circuit}
\end{figure}

The TDM readout architecture has the TES detectors arrayed into a logical row/column grid, with the circuit diagram depicted in Fig \ref{circuit}.  A voltage bias is applied simultaneously to the detectors of an entire column, while the current from each TES is inductively coupled to a unique first stage SQUID amplifier (SQ1).  A bias is applied to a column of SQ1s, while only one is activated at time by using a pair of flux activated switches.  The SQ1s for a column are connected in series with the input to the SQUID Series Array (SSA) amplifier.  The switches are controlled by the Row Select (RS) and Chip Select (CS) signals opening the switches for one row at a time, so that each column conveys a voltage corresponding to a single TES at a time.  Using two switches in parallel with each SQ1 allows us to simultaneously increase the number of rows while reducing the cables and connections.  The SSA amplifier is located at the 4K stage and provides additional cold amplification before the signal is input to the room temperature amplifier in the warm electronics.  Further detail on the CMB-S4 readout scheme can be found in \cite{Barron22}.

In order to achieve the required multiplexing factor of 80, the new system will need to achieve greater bandwidth than previous implementations so that the row switching rate can be fast enough to maintain a sufficient sampling rate.  AdvACT has demonstrated the 64 row readout on a CMB telescope with a row rate of 500 kHz, while BICEP Array has achieved 770 kHz on its latest receiver with 41 rows and X-ray systems have demonstrated row rates of 6.25 MHz with 32 rows \cite{Henderson16, Fatigoni23, Doriese16}.  The model of the readout system presented here is intended to aid in the optimization of the CMB-S4 readout, so that both bandwidth and noise meet the system requirements.

\section{Readout Model}\label{sec2}
We have developed an electrical model that can estimate the readout noise, including the system bandwidth, for a particular configuration of cryogenic amplifiers and room temperature control electronics.  Having an end-to-end model enables the calculation of component-level requirements that will satisfy the system-level specifications.

This model, implemented in Python, works by calculating the noise contributions from the Johnson noise of resistive components and the noise of warm amplifiers for input and output signals at the SSA, SQ1 and TES stages. Each is processed using the gains and bandwidths of the various SQUID and warm amplifiers, including the low-pass filters defined by the resistances, capacitances, and inductances of various circuit components.  By assuming that the gains are all linear, this model is effectively using a small signal approximation, which is applicable to the bolometer output.  The SQUID gains are also used so that all of the noise components can be collectively referenced to a common point in the circuit (generally either the voltage noise at the input to the room temperature amplifier or the current noise at the TES).  Low pass filters are applied to simulate the bandwidth of the signal, with the filters in the room temperature circuits as well as those due to cryogenic traces and wiring.  The overall bandpass of the readout is additionally applied, from those filters on the room temperature amplifiers that filter the overall circuit.  Each individual noise component can then be added in quadrature to find the overall noise of the entire readout system.

While the full bandwidth noise of the readout system is important to properly estimate the design drivers, the end product of the model is to calculate the noise when it is sampled at the multiplexed rate.  While each individual detector is sampled at the multiplexing rate, the bandwidths of the SSA and SQ1 need to extend out to the row rate so that the time for all of the switched inputs to settle at their new values is less than the dwell time spent waiting before recording a measurement and moving on to the following row.  This means that the system bandwidth is significantly larger than the multiplexed sampling rate, and that high frequency noise gets aliased into the sampled band (Fig. \ref{muxing}).

In order to simulate the noise as measured, we generate a random timestream with the modeled noise over the full system bandwidth, and then resample at the multiplexed rate to calculate the aliased noise level.  Due to the fact that 1/f noise from input amplifiers can exceed the aliased white noise level at very low frequencies, the 1/f spectral component is added in quadrature to minimize the computational costs of generating full bandwidth timestreams that are long enough to model the low frequency end of the science band at 0.1 Hz.

To generate the model, inputs for the warm components were used from listed datasheets.  For gains of the cryogenic amplifiers, we have used in situ measurements taken during tuning of the SQUIDs.  For some cryogenic parameters limiting the bandwidth (for example the inductance of the cables and traces making up the SQ1 loop) we have used fits from noise spectra to extract the inductance from the measured bandwidth.  This is a practical method for measuring these parameters, but does mean that those aspects of the model are not tested by their comparison with measured data.

\section{Validation Testing}\label{sec3}
In order to test the results of the model, we have compared it to noise data taken with a Multi-channel Electronics (MCE) box \cite{Battistelli08}.  The MCE is a well understood system that has been used extensively as the warm electronics for TDM readout across many CMB collaborations over the last two decades \cite{Henderson16, Fatigoni23}. To test at different stages, we have taken measurements of the noise on a single pixel with an MCE at the 50 MHz master clock rate, with no biases applied, with the SSA biased but SQ1s superconducting and with the SQ1 biased but TES superconducting, so that we can inspect the different noise components that dominate in each of these configurations.  These tests used a single-level switch SQ1 architecture (NIST MUX15) and lower gain SSAs (NIST SA13) as a crosscheck before applying the model to the new MUX components.  We have also taken data while multiplexing in order to compare the overall noise, as well as downsampled single pixel data that makes it possible to compare intermediate frequencies.

\begin{figure}[htbp]%
\centering
\includegraphics[width=0.8\textwidth]{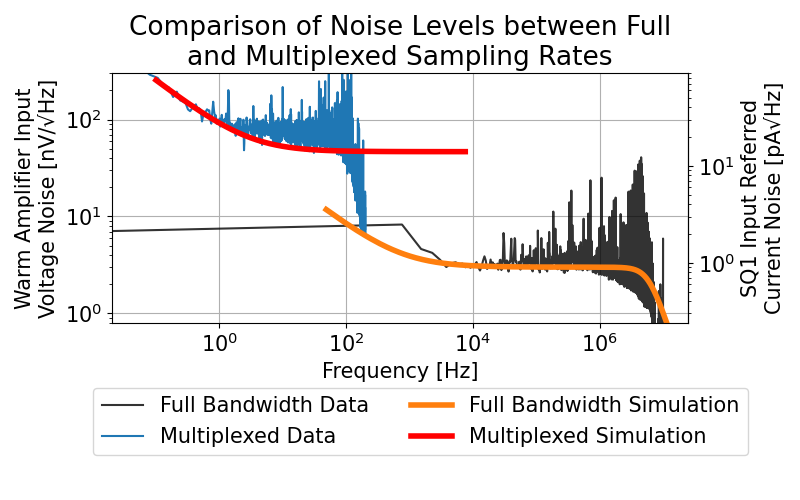}
\caption{Comparison of noise levels measured up to the full bandwidth of the MCE with noise measured while multiplexing.  This agreement in white noise levels between the simulation and measurement validate the aliasing calculation preformed by the model.  The 1/f component visible on both plots represents the noise of the room temperature amplifier, and is one of the components whose specifications must be set by this modeling work.}\label{muxing}
\end{figure}
\begin{figure}[htbp]%
\centering
\includegraphics[width=0.95\textwidth]{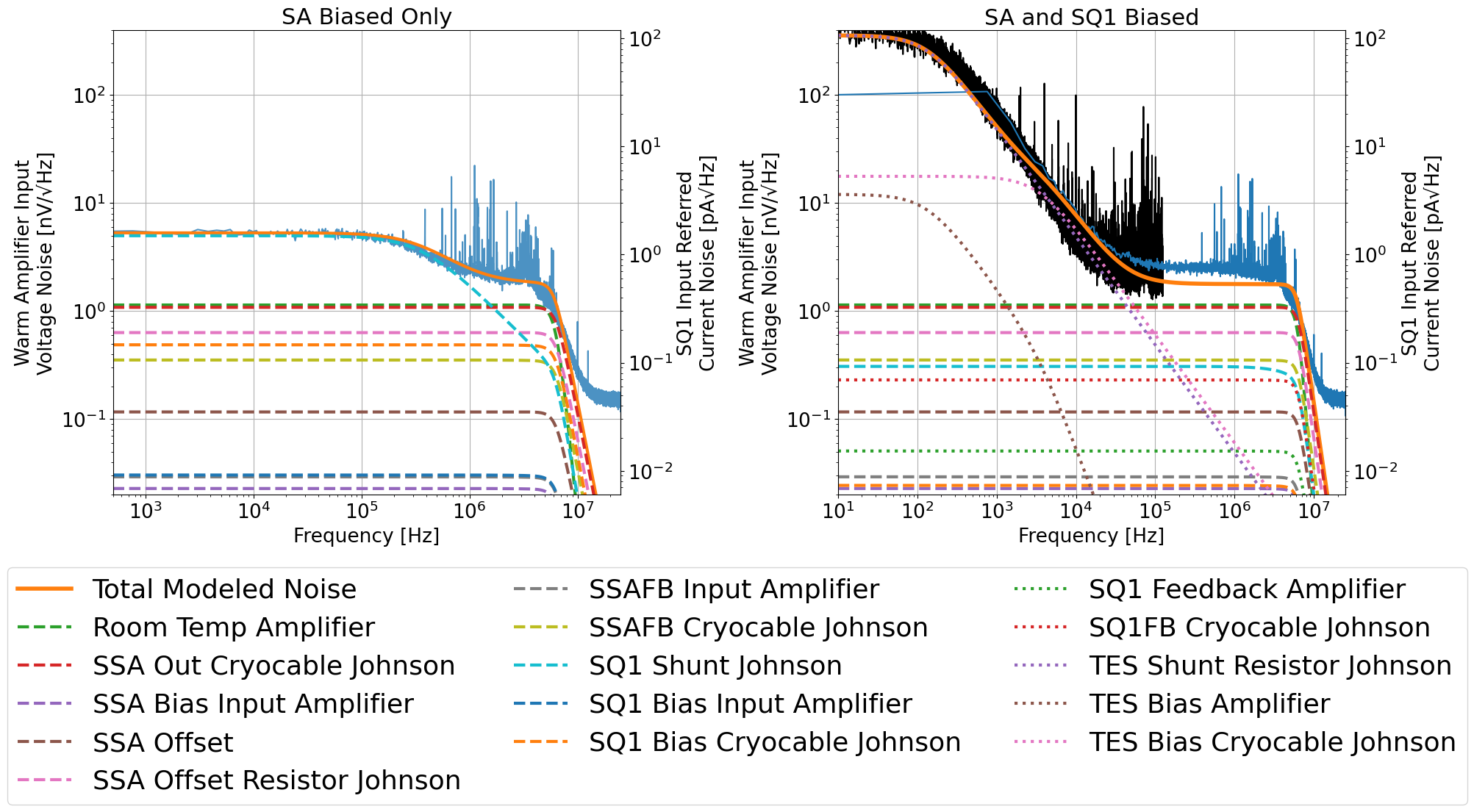}
\caption{Left: Comparison of noise level measured up to the full bandwidth of the MCE with the expectations of the model, with SSA biased but the SQ1s left superconducting and row select switches all closed.  The Johnson noise of the SQ1 shunt resistor (labeled 1 $\Omega$ in Fig. \ref{circuit}) dominates the noise spectrum, since all of the other components in the SQ1 loop are superconducting.  Note that while the noise level serves to validate the model, the inductance that limits the bandwidth is calculated based on these noise measurements and is therefore not an independent test.
Right: Comparison of the noise measured with an MCE to the expectations of the model, with SSA and SQ1 biased and one row select switch open.  The noise was measured in open loop out to the full bandwidth of the MCE (blue) as well as at a reduced rate in closed loop (black) in order to sample lower frequencies.  The Johnson noise of the TES shunt resistor dominates the low frequency noise spectrum, since the TES is superconducting for this measurement, while the bandwidth is limited by the Nyquist inductor.  High frequency lines in the measured response are due to pickup from the environment and show up at other positions in the lower rate data due to aliasing.  The disagreement between model and measurement in the 100 kHz to 1 MHz range is also connected to environmental pickup based on comparison to multiple measurements.}\label{model}
\end{figure}

Comparisons between the model and measurement for these scenarios can be found in Figures \ref{muxing} and \ref{model}.  Comparing the multiplexed and full bandwidth measurements to the model (Fig. \ref{muxing}), we see that the noise spectra match the model predictions, validating the calculation we have done of aliasing due to multiplexed sampling.  The measured noise spectra feature a variety of MHz frequency lines, which are not present in the model.  These are due to pickup from the cryostat environment and are not surprising, given known nonidealities in our setup, which will be addressed in future work.  It is also worth noting that while the broadband portions of the unbiased and SSA biased spectra are fairly aligned with the model, there is a noticeable disagreement with the SQ1 biased data in the 100 kHz--1 MHz range.  Based on day-to-day variations in this disagreement as well as differences between different readout channels, the current best explanation for this is cryogenic common-mode pickup in the SQ1 loop circuit.  To improve on this, CMB-S4 is looking into the use of differential cryogenic readout which is under development \cite{Durkin23}.  Otherwise, the model tends to agree with measurement, although bandwidth limiting inputs that are measured from noise spectra, such as the $\sim$200 kHz rolloff of the SQ1 shunt Johnson noise in Fig. \ref{model}, cannot be considered validated since the model inputs are taken from that same test.

\section{Model Application to Design}\label{sec4}
As we develop the prototypes of the CMB-S4 room temperature electronics, we can compare their performance to the predictions of the model.  Measurements of the full bandwidth noise have shown the expected level of amplifier noise when testing on a benchtop for different choices of amplifier.  We are testing a variety of warm readout circuits, including the use of dedicated ASICs \cite{Gonzalez22} and the incorporation of differential readout.

Another important application of this model is setting the component level requirements for the amplifiers that will populate the new readout.  As can be seen in the model spectra (Fig. \ref{model}), the overall noise level is very sensitive to the noise of the room temperature amplifier that amplifies the SSA output, so it is important to choose the lowest noise amplifier possible, while other amplifiers contribute nearly two orders of magnitude less noise, so some more flexibility is possible.  Amplifier choices will ensure that the warm electronics noise will be within the overall noise budget for the CMB-S4 readout, and we already have amplifiers in hand that will be sufficient in that regard.

One takeaway from the model comparison with MCE data is that the Johnson noise of the manganin cryocable that connects the 4K SSA with the 300K electronics can be a dominant contributor to the overall noise, given reasonable choices of warm amplifiers.  While the high thermal resistance of this cable is important for maintaining cryogenic isolation of the 50K and 4K stages of the cryostat, this model has motivated the possibility of a hybrid cable that would have lower impedance on the SSA wire alone.  Such a cable would allow us to improve the noise without needing to compromise the thermal impedance of most of the cable.

These noise measurements have also demonstrated the ability to use the Johnson noise of the SQ1 shunt resistor to evaluate the inductance of the SQ1 loop.  The SQ1 loop bandwidth has been shown to be among the limiting factors to the overall bandwidth \cite{Durkin21}, so minimizing this inductance will enable us to cycle faster and thus increase the multiplexing rate.  By repeating this measurement while excluding the SQ1s themselves, we can measure the inductance of the cables connecting the 4K and 100 mK stages and set requirements on it if necessary to open increased bandwidth.

\section{Conclusions}\label{sec5}
We have developed a model of the CMB-S4 readout system that calculates the noise and bandwidth for a given set of TDM readout parameters.  This model has been compared to data taken with an MCE and is well aligned with the measurements on legacy readout components, validating the model's outputs.  By comparing to the dominant noise sources in those results, we have identified potential upgrades to improve the system performance.   Moving forwards, the model will be critical in developing component-level requirements and performing quality assurance as we test the new CMB-S4 electronics.
\backmatter

\bmhead{Acknowledgments}
CMB-S4 is supported by the U.S. Department of Energy (DOE), Office of High Energy Physics (HEP) under Contract No. DE–AC02–05CH11231; by the National Energy Research Scientific Computing Center, a DOE Office of Science User Facility under the same contract; and by the Divisions of Physics and Astronomical Sciences and the Office of Polar Programs of the U.S. National Science Foundation under Mid-Scale Research Infrastructure award OPP-1935892. Work at SLAC National Accelerator Laboratory was supported by DOE HEP under contract DE-AC02-76SF00515. D.R.B. is supported by DOE HEP under award number DE-SC0021435, and NSF’s Office of Integrative Activities under award OIA-2033199. J.P.F. is supported by DOE HEP under award number DE-SC0015655.  Considerable additional support is provided by the many CMB-S4 collaboration members and their institutions.

\bibliography{sn-bibliography}

\end{document}